\documentclass[
showpacs
]{revtex4}
\usepackage{makeidx}
\usepackage{amssymb}
\usepackage{amsmath}
\usepackage{mathrsfs}
\usepackage{graphicx}
\usepackage{dcolumn}
\usepackage{bm}
\usepackage[center]{subfigure}
\usepackage{color}

\begin{document}

\title{Gap solitons in Rabi lattices}
\author{Zhaopin Chen$^{1}$}
\author{Boris A. Malomed$^{1,2}$}
\affiliation{$^{1}$Department of Physical Electronics, School of Electrical Engineering,
Faculty of Engineering, Tel Aviv University, Tel Aviv 69978, Israel\\
$^{2}$Laboratory of Nonlinear-Optical Informatics, ITMO University, St.
Petersburg 197101, Russia}

\begin{abstract}
We introduce a two-component one-dimensional system, which is based on two
nonlinear Schr\"{o}dinger/Gross-Pitaevskii equations (GPEs) with spatially
periodic modulation of linear coupling (``Rabi lattice") and self-repulsive
nonlinearity. The system may be realized in a binary Bose-Einstein
condensate, whose components are resonantly coupled by a standing optical
wave, as well as in terms of the bimodal light propagation in periodically
twisted fibers. The system supports various types of gap solitons (GSs),
which are constructed, and their stability is investigated, in the first two
finite bandgaps of the underlying spectrum. These include on- and
off-site-centered solitons (the GSs of the off-site type are additionally
categorized as spatially even and odd ones), which may be symmetric or
antisymmetric, with respect to the coupled components. The GSs are chiefly
stable in the first finite bandgap, and unstable in the second one. In
addition to that, there are narrow regions near the right edge of the first
bandgap, and in the second one, which feature intricate alternation of
stability and instability. Unstable solitons evolve into robust breathers or
spatially confined turbulent modes. On-site-centered GSs are also considered
in a version of the system which is made asymmetric by the Zeeman effect, or
by birefringence of the optical fiber. A region of alternate stability is
found in the latter case too. In the limit of strong asymmetry, GSs are
obtained in a semi-analytical approximation, which reduces two coupled GPEs
to a single one with an effective lattice potential.
\end{abstract}

\pacs{42.65.Tg; 42.70.Qs; 05.45.Yv}
\maketitle

%\email{malomed@post.tau.ac.il}

%\author{Shenhe Fu$^{2}$}
%\author{Yongyao Li$^{3}$}

%$^{2}$State Key Laboratory of Optoelectronic Materials and Technologies,\\
%Sun Yat-sen University, Guangzhou 510275, China\\
%$^{3}$Department of Applied Physics, South China
%Agricultural University, Guangzhou 510642, China.}

\section{Introduction}

Solitons in lattice potentials have drawn a great deal of interest in recent
decades, as they occur in diverse physical settings, and exist in many
different varieties \cite{early-reviews}-\cite{Kartashov}. Periodic
potentials, added to the underlying nonlinear Schr\"{o}dinger or
Gross-Pitaevskii equations (GPEs), help to create and stabilize solitons
which do not exist or are unstable in free space. In optics, one source of
the spatial periodicity is provided by Bragg gratings \cite{Bragg}.
Effective periodic potentials can also be readily induced by photonic
crystals, which can be built as permanent structures by means of various
techniques \cite{Lederer}-\cite{Szameit}, or as virtual lattices formed by
interfering laser beam in photorefractive crystals \cite{Efremidis,Lederer}.
For matter waves in atomic Bose-Einstein condensates (BECs), perfect
periodic potentials are imposed by optical lattices, i.e., interference
patterns constructed by counter-propagating coherent optical beams \cite%
{Pitaevskii,early-reviews}. In the presence of the self-repulsive
nonlinearity, localized modes which self-trap in periodic potentials are
usually called \textit{gap solitons} (GSs), as they exist in bandgaps of the
underlying Bloch spectrum induced by the potential in the linear
approximation \cite{Wu}-\cite{HS}. Different families of GSs are
distinguished by the number of the bandgaps in which they reside.

In the presence of an appropriate periodic potential, binary BECs \cite%
{Myatt} and binary photonic systems \cite{Mak,Skryabin} can host
two-component GSs, which have been theoretically elaborated in various
settings \cite{Mak}-\cite{Athikom}. The use of the Feshbach-resonance
technique \cite{Simoni}, that switches the repulsion between atoms into
attraction, makes it possible to create two-component \textit{symbiotic
solitons} in binary BEC, which are supported by attraction between the two
components, while each of them is subject to self-repulsion \cite{Perez}.
This concept was extended to symbiotic GSs in a system of two mutually
repelling components loaded into a common lattice potential \cite{Athikom}.
Another extension of this concept was elaborated for dark solitons in spinor
systems \cite{dark}.

An essential ingredient of many two-component systems is linear
interconversion (Rabi coupling) between the components. In binary BEC, the
interconversion is driven by a resonant electromagnetic field, which couples
different atomic states representing the components \cite{Ballagh,Rabi}.
Two-component GSs coupled by the linear interconversion were studied too
\cite{Adhikari,Yongyao1}. In optics, the Rabi coupling is emulated by the
linear coupling between copropagating waves in dual-core waveguides \cite%
{coupler}. In particular, GSs in a dual-core Bragg grating were studied in
Ref. \cite{Mak}. Similarly, a dual-core BEC trap may hold two matter-wave
fields with an effective Rabi coupling between them, provided by tunneling
of atoms across a gap separating the two cores \cite{BEC-coupler}. In this
connection, it is relevant to mention that GSs may be supported by \textit{%
Zeeman lattices}, i.e., spatially periodic modulation of the difference in
the chemical potential between two BEC\ components which are linearly
coupled by spin-orbit coupling \cite{Zlattice}.

In the present work, our objective is to propose a different mechanism of
the creation of two-component GSs, without the use of any lattice potential,
but rather making use of the linear coupling between two components of the
wave field in the form of a standing wave, which may be called a ``Rabi
lattice" (cf. ``Rabi management", i.e., the linear coupling with a
time-periodic coefficient, introduced in Ref. \cite{Rabi}), assuming
intrinsic self-repulsion in each component. This setting may be realized in
a binary BEC illuminated by a pair of counterpropagating resonantly-coupling
waves, the interference of which builds the standing wave. In this
connection, it is relevant to mention recent work \cite{Harel}, in which it
was demonstrated that a spatially localized (rather than periodically
modulated) linear coupling between the components may play a role in the
soliton dynamics similar to that of localized attractive potentials. We here
focus on the basic case of the linear coupling, neglecting nonlinear
cross-repulsion between the components (it can be suppressed by means of the
Feshbach resonance \cite{cross-F}). An extended version of the system
includes asymmetry between the two components, which may be imposed by the
Zeeman splitting between them. As shown in the next section, a similar model
can be implemented in nonlinear optics, considering the co-propagation of
two polarizations in a periodically twisted fibers, while the asymmetry may
be imposed by the fiber's intrinsic birefringence.

In the symmetric system, the equations for the two components merge into a
single one, if solutions with equal components are looked for. In this case,
the standing-wave-shaped linear coupling turns into an effective lattice
potential. Although shapes of the corresponding GSs are known, a new problem
is their stability in the framework of the two-component system, as well as
constructing GSs in the asymmetric one. In this work, we concentrate on the
1D setting, while the 2D version, which is possible in BEC, will be
considered elsewhere.

The rest of the paper is structured as follows. In Sec. II, we introduce the
model and present a method for the study of stability of the GSs. Several
families of GSs and results for their stability are produced in Sec. III.
The solitons are classified as on- and off-site-centered ones and symmetric
or antisymmetric, with respect to the two component. The off-site-centered
GSs are additionally categorized as spatially even or odd modes. In Sec. IV,
we address on-site-centered GSs in an asymmetric version of the system, with
the objective to identify their existence and stability areas. For a
strongly asymmetric system, analytical approximation is developed too. The
paper is concluded by Sec. V.

\section{The model}

The two-component system with the linear-coupling coefficient spatially
shaped as the standing wave (Rabi lattice) is based on the system of scaled
GPEs for two components of the mean-field wave function, $u(x,t)$ and $%
v(x,t) $:%
\begin{eqnarray}
\left( i{\frac{\partial }{\partial t}}+{\frac{1}{2}}{\frac{\partial ^{2}}{%
\partial x^{2}}}-\sigma |u|^{2}\right) u+\epsilon \cos (2x)\cdot v+bu &=&0,
\label{basicEqU} \\
\left( i{\frac{\partial }{\partial t}}+{\frac{1}{2}}{\frac{\partial ^{2}}{%
\partial x^{2}}}-\sigma |v|^{2}\right) v+\epsilon \cos (2x)\cdot u-bv &=&0.
\label{basicEqV}
\end{eqnarray}%
where $\epsilon $ is the amplitude of the Rabi lattice, whose period is
fixed to be $\pi $ by making use of obvious rescaling, and real coefficient $%
b$ accounts for possible asymmetry introduced by the Zeeman splitting.
Results are reported below for $\epsilon =6$ in Eqs. (\ref{basicEqU}) and (%
\ref{basicEqV}), which adequately represents the generic situation.

Further, $\sigma =+1$ and $-1$ correspond to the self-defocusing and
focusing signs of the contact nonlinearity. Being interested in GSs, we
focus on the case of self-defocusing, $\sigma =+1$, which cannot create
regular solitons in free space. Effects of the nonlinear interaction between
the two components, with relative strength $g$ (the use of the Feshbach
resonance makes it relevant to consider all the cases of $g>0$, $g=0$, and $%
g<0$ \cite{cross-F}), are accounted for by the addition of
cross-phase-modulation (XPM) terms to the nonlinearity in these equations:%
\begin{equation}
\left\vert u\right\vert ^{2}u\rightarrow \left( |u|^{2}+g|v|^{2}\right)
u,~|v|^{2}v\rightarrow \left( |v|^{2}+g|u|^{2}\right) v.  \label{g}
\end{equation}%
We chiefly disregard the XPM terms here (as mentioned above, in BEC\ this
interaction may be eliminated with the help of the Feshbach-resonance
method), except for the limit case of strong asymmetry [large $b$ in Eqs. (%
\ref{basicEqU}) and (\ref{basicEqV})], in which the XPM can be easily taken
into account in the framework of the semi-analytical approximation, see Eqs.
(\ref{vu})-(\ref{delta_k}) below. In a systematic form, XPM effects will be
considered elsewhere.

Unlike the BEC system, introduction of a similar model in terms of optical
dual-core waveguides is problematic, as the coefficient of the inter-core
coupling cannot, normally, change its sign. On the other hand, the same
linear coupling as defined in Eqs. (\ref{basicEqU}) and (\ref{basicEqV}) may
naturally appear in the model of the co-propagation of two linear
polarizations of light in a ``rocking" optical fiber, subject to a
periodically modulated twist \cite{rocking} (with evolution variable $t$
replaced by the propagation distance, $z$, and $x$ replaced by the reduced
temporal variable, $\tau $), while parameter $b$ represents the
phase-velocity birefringence of the fiber. In the latter case, however, the
XPM terms with $g=2/3$ in Eq. (\ref{g}) should be taken into account,
therefore this case too will be considered in detail elsewhere. In
principle, more freedom in the choice of the XPM coefficient is offered by
photonic-crystal fibers, which may also carry periodic twist \cite{PCF}.
Lastly, the optical model may include a group-velocity birefringence too,
accounted for by additional terms $\symbol{126}\left( +u_{\tau },-v_{\tau
}\right) $ in the respective equations, although this effect is usually much
weaker than the phase-velocity birefringence.

\begin{figure}[tbp]
\centering{\label{fig1a} \includegraphics[scale=0.25]{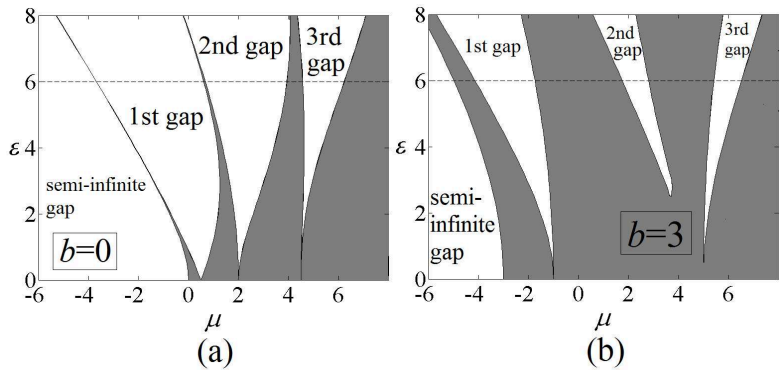}}
\caption{(Color online) The bandgap spectrum of the linearized version of
Eqs. (\protect\ref{basicEqU}) and (\protect\ref{basicEqV}) with $\protect%
\epsilon =6$, for $b=0$ (a) and $b=3$ (b), which correspond, respectively,
to the symmetric and asymmetric system. Shaded areas designate Bloch bands
where gap solitons do not exist. The semi-infinite gap and lowest finite
ones are labeled. }
\label{blochband}
\end{figure}

Stationary solutions of Eqs. (\ref{basicEqU}) and (\ref{basicEqV}) are
looked for as usual,
\begin{equation}
u(x,t)=e^{-i\mu t}U(x),v(x,t)=e^{-i\mu t}V(x),  \label{statSol}
\end{equation}%
where $\mu $ is a real chemical potential, and real wave functions $U$ and $%
V $ obey the stationary equations, as said above:
\begin{eqnarray}
\mu U+\frac{1}{2}U^{\prime \prime }-U^{3}+\epsilon \cos (2x)\cdot V+bU &=&0,
\label{statEqsU} \\
\mu V+\frac{1}{2}V^{\prime \prime }-V^{3}+\epsilon \cos (2x)\cdot U-bV &=&0,
\label{statEqsV}
\end{eqnarray}%
where the prime stands for $d/dx$ (hereafter, we fix the defocusing sign of
the nonlinearity, $\sigma =+1$, as said above). Numerical solutions of Eqs. (%
\ref{statEqsU}) and (\ref{statEqsV}) was produced by means of the Newton's
method. The bandgap spectrum generated by the solution of the linearized
version of the equations is shown in Fig. \ref{blochband}.

Equations (\ref{basicEqU}) and (\ref{basicEqV}) conserve the total norm,
i.e., scaled number of atoms, in terms of BEC,
\begin{equation}
N=\int_{-\infty }^{+\infty }\left( |u|^{2}+|v|^{2}\right) dx\equiv
N_{u}+N_{v}~,  \label{P}
\end{equation}%
%
%
%
%
%
%{\LARGE [due to the change in the notation, }${\LARGE P}${\LARGE \ in all
%figures should be replaced by }${\LARGE N}${\LARGE ]}
and the Hamiltonian,
\begin{gather}
H=\int_{-\infty }^{+\infty }\left[ \frac{1}{2}\left( \left\vert \frac{%
\partial u}{\partial x}\right\vert ^{2}+\left\vert \frac{\partial v}{%
\partial x}\right\vert ^{2}+|u|^{4}+|v|^{4}\right) \right.  \notag \\
\left. -\epsilon \left( u^{\ast }v+uv^{\ast }\right) \cos \left( 2x\right)
+b\left( |v|^{2}-|u|^{2}\right) \right] dx~.  \label{H}
\end{gather}%
where $\ast $ stands for the complex conjugate, while the conservation of
the momentum is destroyed by the presence of the Rabi lattice. Asymmetry of
two-component solitons is determined by the respective ratio,
\begin{equation}
R=(N_{u}-N_{v})/(N_{u}+N_{v}).  \label{R}
\end{equation}%
For families of GSs, $N$ and $H$, as well as $R$ (if $R\neq 0$), may be
naturally considered as functions of chemical potential $\mu $.

Stability of stationary solutions can be investigated by means of the
linearization against small complex perturbations $\phi _{1,2}(x,z)$ and $%
\psi _{1,2}(x,z)$ added to solution (\ref{statSol}):

\begin{eqnarray}
u(x,t) &=&e^{-i\mu t}[U(x)+\phi _{1}(x)e^{-i\lambda t}+\phi _{2}^{\ast
}(x)e^{i\lambda ^{\ast }t}],  \notag \\
v(x,t) &=&e^{-i\mu t}[V(x)+\psi _{1}(x)e^{-i\lambda t}+\psi _{2}^{\ast
}(x)e^{i\lambda ^{\ast }t}],  \label{perturbEq}
\end{eqnarray}%
where $\lambda $ is the perturbation eigenfrequency, which may be complex.
Instability takes place if there is at least one eigenvalue with Im$(\lambda
)>0$. Oscillatory instabilities correspond to complex $\lambda $, with both
real and imaginary parts different from zero, which happens in previously
studied related systems \cite{Louis,Athikom}. The substitution of
expressions (\ref{perturbEq}) into Eqs. (\ref{basicEqU}), (\ref{basicEqV})
and subsequent linearization leads to the eigenvalue problem for $\lambda $,
based on the following system of equations:

\begin{eqnarray}
-{\frac{1}{2}}\phi _{1}^{\prime \prime } &+&U^{2}\left( 2\phi _{1}+\phi
_{2}\right) -\left( b+\mu \right) \phi _{1}-\epsilon \cos (2x)\psi _{1}
\notag \\
&=&\lambda \phi _{1},  \notag \\
{\frac{1}{2}}\phi _{2}^{\prime \prime } &-&U^{2}\left( 2\phi _{2}+\phi
_{1}\right) +\left( b+\mu \right) \phi _{2}+\epsilon \cos (2x)\psi _{2}
\notag \\
&=&\lambda \phi _{2},  \notag \\
-{\frac{1}{2}}\psi _{1}^{\prime \prime } &+&V^{2}\left( 2\psi _{1}+\psi
_{2}\right) +\left( b-\mu \right) \psi _{1}-\epsilon \cos (2x)\phi _{1}
\notag \\
&=&\lambda \psi _{1},  \notag \\
{\frac{1}{2}}\psi _{2}^{\prime \prime } &-&V^{2}\left( 2\psi _{2}+\psi
_{1}\right) -\left( b-\mu \right) \psi _{2}+\epsilon \cos (2x)\phi _{2}
\notag \\
&=&\lambda \psi _{2}~.  \label{eigenEq}
\end{eqnarray}%
These equations can be rewritten in the matrix form, $\hat{M}(\phi _{1},\phi
_{2},\psi _{1},\psi _{2})^{T}=\lambda (\phi _{1},\phi _{2},\psi _{1},\psi
_{2})^{T}$, where operator $\hat{M}$ corresponds to the matrix in the
left-hand side of Eqs. (\ref{eigenEq}). For the numerical solution of the
stability problem, we discretize functional expressions in matrix elements
by means of the center-difference numerical scheme, and then calculate the
eigenvalue spectrum of the matrix, truncated to a sufficiently large finite
size, for stationary solutions.

\section{Gap solitons}

The numerical solution of the symmetric version of Eqs. (\ref{statEqsU}) and
(\ref{statEqsV}), with $b=0$, shows that the system produces several basic
types of GSs. With respect to their spatial structure, they can be
classified as on- and off-site-centered localized modes, which feature,
respectively, a single density maximum coinciding with a local minimum of
the modulation function, $-\epsilon \cos (2x)$, i.e., $x=0$, or two density
maxima placed at adjacent modulation maxima, $x=\pm \pi /2$. Further, the
off-cite-centered modes, with the pair of density maxima, may be spatially
even or odd as functions of $x$ (all onsite-centered modes are, obviously,
of the even type). Then, in the absence of the Zeeman splitting ($b=0$), the
two-component GSs are categorized as ``symmetric" or ``antisymmetric", if
their two components are, respectively, identical, or differ by opposite
signs. In the following analysis, we first focus on on-site-centered
symmetric, off-site-centered even antisymmetric, and off-sited odd
antisymmetric GS species. We also consider on-site-centered antisymmetric,
off-site-centered even symmetric, and off-site-centered odd symmetric modes,
whose stability is, severally, the same as that of the three above-mentioned
species. The analysis is performed for the GSs residing in the first and
second finite bandgaps, see Fig. \ref{blochband}.

\subsection{On-site-centered symmetric gap solitons}

\begin{figure}[tbp]
\centering{\label{fig2a} \includegraphics[scale=0.2]{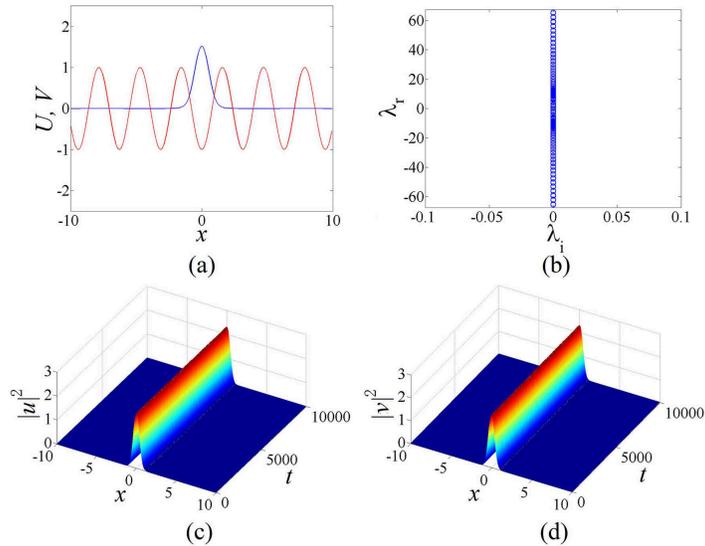}}
\caption{(Color online) (a) A typical example of stable on-site-centered
symmetric GSs, found in the first finite bandgap, at $\protect\mu =-2$, $b=0$%
, and $\protect\epsilon =6$. Here, and in similar figures below, the
background pattern (red sinusoid) represents the scaled underlying Rabi
lattice [periodic modulation of the coupling constant, $-\left( \protect%
\epsilon /6\right) \cos \left( 2x\right) $, with scaling factor $1/6$ added
to keep the sinusoid within boundaries of the panel]. (b) The eigenvalue
spectrum for small perturbations around the soliton, which confirms its
stability. It is further corroborated by simulations of the evolution of the
soliton initially perturbed by random noise at the level of $5\%$ of the
soliton's amplitude, which are displayed in panels (c) and (d).}
\label{systable}
\end{figure}

\begin{figure}[tbp]
\centering{\label{fig3a} \includegraphics[scale=0.2]{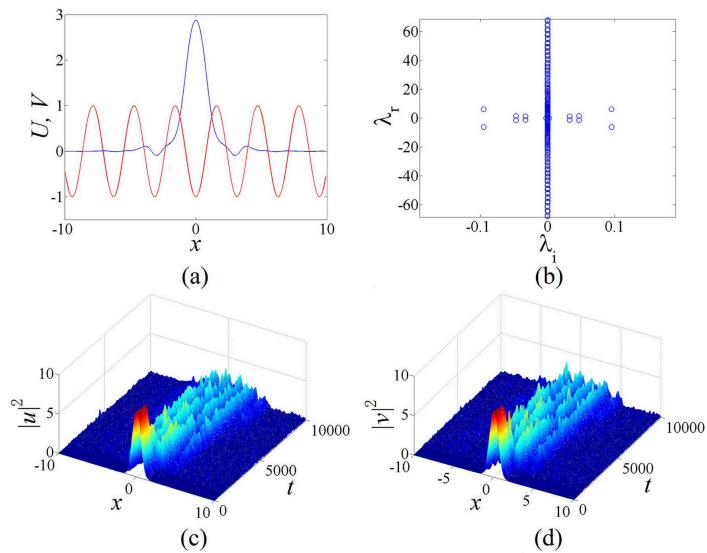}}
\caption{(Color online) The same as in Fig. \protect\ref{systable}, but for
a typical unstable on-site-centered symmetric soliton in the second finite
bandgap, at $\protect\mu =3$ and $\protect\epsilon =6$. As seen in panels
(c) and (d), development of the oscillatory instability replaces the GS by a
``turbulent" pattern, which remains spatially confined.}
\label{syunstable}
\end{figure}

\begin{figure}[tbp]
\centering{\label{fig4a} \includegraphics[scale=0.2]{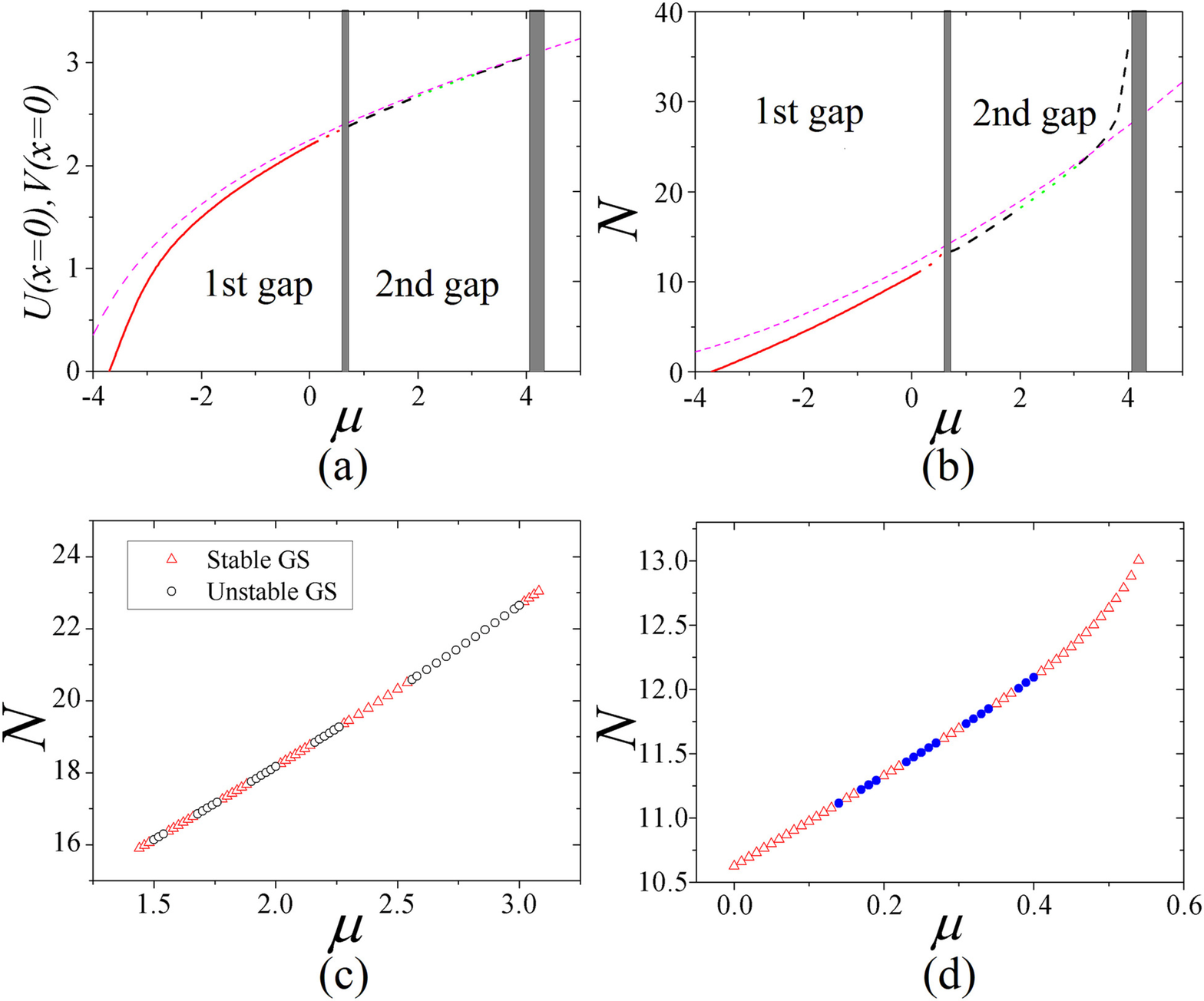}}
\caption{(Color online) (a) The numerically found amplitude of fundamental
on-site-centered symmetric GSs versus chemical potential $\protect\mu $, for
$\protect\epsilon =6$ and $b=0$. The dashed magenta curve is produced by the
Thomas-Fermi approximation (TFA) for the single-component GSs: $%
U(x=0)=V(x=0)=\protect\sqrt{6+\protect\mu }-3(6+\protect\mu )^{-3/2}$. Panel
(b) shows the total norm $N$ versus $\protect\mu $, along with its
dashed-magenta TFA counterpart, $N=2\left[ \protect\sqrt{36-\protect\mu ^{2}}%
+\protect\mu \cos ^{-1}(-\protect\mu /6)\right] $. In these panels, as well
as in Fig. \protect\ref{offsite:stability}(a) below, red solid and black
dashed segments represent stable and unstable GSs, respectively, while the
green dotted segment designates a region of alternate stability and
instability. The red dotted segment near the right edge of the first bandgap
designates a region of alternating stable GSs and ones weakly unstable
against oscillatory perturbations. (c) The alternation of the stability and
instability in the green dotted segment of (b). Here, red triangles and
black circles represents stable and unstable GSs, respectively. (d) The
alternation of the stability and weak oscillatory instability in the red
dotted segment of (b). Here, and also in Fig. \protect\ref{offsite:stability}%
(c) below, red triangles and blue filled circles represent stable GSs and
ones subject to the weak instability, respectively.}
\label{Syrelationship}
\end{figure}

It is obvious that on-site-centered symmetric GSs [see typical examples in
Figs. \ref{systable}(a,b) and \ref{syunstable}(a,b)], with equal components,
$U(x)=V(x)$ (in the system with $b=0$), are identical, in their shape, to
the GSs produced by the single-component GPE, the Rabi lattice becoming
equivalent to a single-component lattice potential, $-\epsilon \cos \left(
2x\right) $. Accordingly, the density maximum placed at $x=0$ in Figs. \ref%
{systable}(a) and \ref{syunstable}(a) tends to minimize the effective
potential. However, the difference in the stability between the
single-component model and the present two-component system is essential, as
shown in Fig. \ref{Syrelationship}(a,b). Recall that the GS family tends to
be stable in the first two finite bandgaps of the single-component model,
except for a weak oscillatory instability, caused by the appearance of
complex eigenvalues, close to the right edge of the first bandgap, and in
the second one \cite{Alfimov}. In addition, dashed magenta lines display, in
both panels (a) and (b)\ of Fig. \ref{Syrelationship}, analytical
predictions based on the Thomas-Fermi approximation (see details in the
figure caption), which was elaborated for single-component GSs in Ref. \cite%
{Athikom}

%{\LARGE [Please replace panels (a) and (b) in Fig. \ref{Syrelationship} by
%ones (which you have now showed in a pdf picture) which include the TFA
%predictions shown by the dashed magenta lines. We do not need to include
%more accurate empiric fits, as they do not provide any essential
%information, while the comparison to the TFA is essential, even if it seems
%less accurate. Also, in panel (a) please show stable and unstable segments
%of the GS family, using exactly the same types of lines as in (b)
%(otherwise, this plot does not display any new results). Please carefully
%check the current caption to the figure, where I have made additional
%changes.]}

\begin{figure}[tbp]
\centering{\label{fig5a} \includegraphics[scale=0.2]{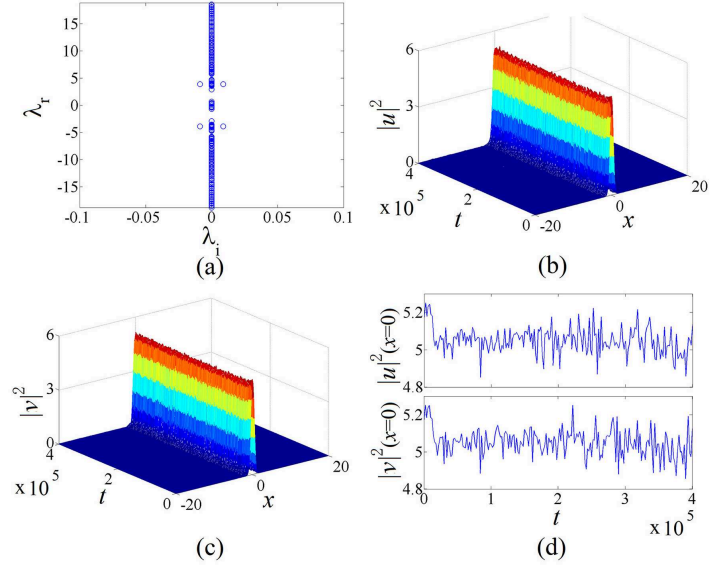}}
\caption{(Color online) An example of a weakly unstable GS with $\protect\mu %
=0.32$, belonging to the red dotted segment near the right edge of the first
finite bandgap in Fig. \protect\ref{Syrelationship}. (a) Unstable
eigenvalues of small perturbations are complex with small imaginary parts.
(b,c) Direct simulations corroborate the weak oscillatory instability of the
soliton, which keeps its localized shape. (d) The weak instability is
additionally illustrated by the time dependence of squared amplitudes of
both components of the same solutions.}
\label{onsyosci}
\end{figure}

Figures \ref{Syrelationship}(a,b) demonstrate that the on-site-centered GSs
are stable in the first finite bandgap of the two-component system (see a
typical example in Fig. \ref{systable}), except for a small region near the
right edge of the bandgap, which roughly resembles the above-mentioned weak
oscillatory instability of single-component GSs near the edge of the first
bandgap \cite{Alfimov}. In this small region, stable solitons alternate with
unstable ones, which are subject to a weak oscillatory instability caused by
complex eigenvalues. A detailed structure of this region is displayed in
Fig. \ref{Syrelationship}(d), and a typical example of a weakly unstable GS,
which keeps a nearly undisturbed shape, is presented in Fig. \ref{onsyosci}.
For $\epsilon =6$, the stable part of the first finite bandgap is $-3.70<\mu
<0.14$ and $0<N<11.11$ in terms of the chemical potential and total norm,
respectively.

In the present system the GSs are primarily unstable in the second finite
bandgap, as is indicated in Fig. \ref{Syrelationship}(b), and illustrated by
a typical example in Fig. \ref{syunstable}, which shows that the unstable
solitons evolve into \emph{spatially confined} chaotic modes (``solitons of
conservative turbulence"). An exception is the green dotted segment, which
contains alternating stable and unstable solitons, as shown in detail in
Fig. \ref{Syrelationship}(c).

%What should be noticed is that a few of stable on-site symmetric GSs in the alternative region(see Fig.1(c)) of stable and unstable on-site symmetric GSs, which are identified by eigenvalues analysis, could also transform into a confined state. Its typical example is displayed in Fig.4 This phenomenon also happens in the case of asymmetric on-site gap solitons. A part of eigenvalue-showing stable GSs in the gray areas of Fig.13 finally become unstable in the process of propagation simulation, with its energy in the mean peak splitting to the other peaks. This phenomena may be caused by nonlinear instability.

\subsection{Off-site-centered spatially-even antisymmetric gap solitons}

\begin{figure}[tbp]
\centering{\label{fig6a} \includegraphics[scale=0.2]{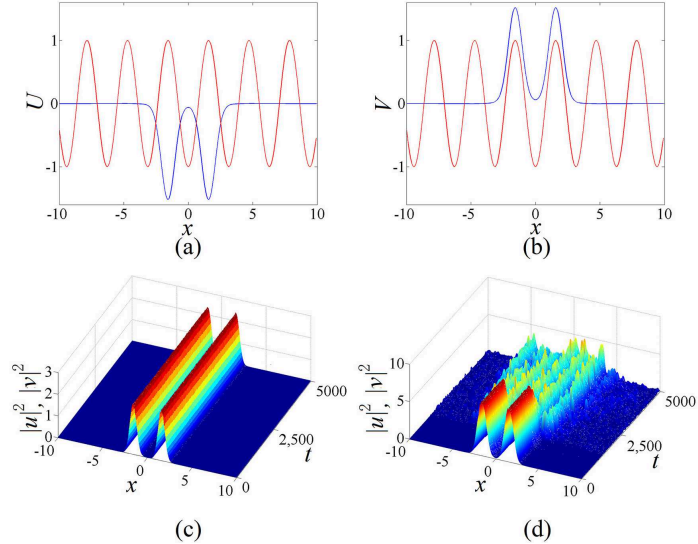}}
\caption{(Color online) Panels (a) and (b) display a typical stable
off-site-centered antisymmetric spatially-even GS for $\protect\mu =-2$,
which falls into the first finite bandgap. (c) Direct simulations (with the
initial random-amplitude perturbation at the $5\%$ level) of the evolution
of the same soliton, which corroborate its stability. (d) Simulations of the
perturbed evolution of the GS of the same type, but belonging to the second
finite bandgap, with $\protect\mu =2$, demonstrate that this unstable
soliton is finally transformed into a spatially confined turbulent state.}
\label{offevenanti}
\end{figure}

\begin{figure}[tbp]
\centering{\label{fig7a} \includegraphics[scale=0.2]{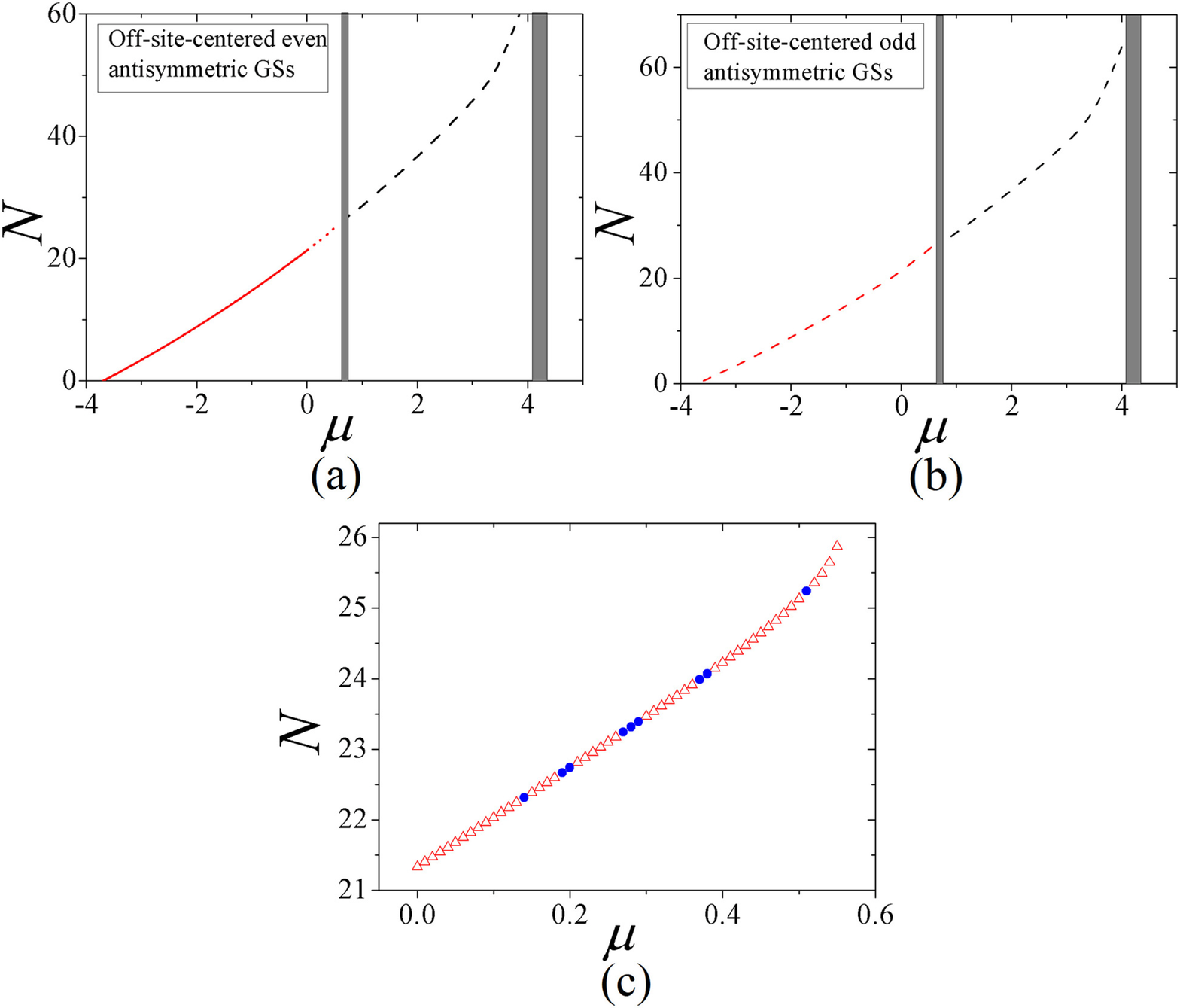}}
\caption{(Color online) (a) Total norm $N$ versus chemical potential $%
\protect\mu $ for off-site-centered antisymmetric spatially-even GSs, cf.
Fig. \protect\ref{Syrelationship}(b) for the on-site centered symmetric
solitons. (b) The $N(\protect\mu )$ curve for the family of off-site-centered
antisymmetric spatially-odd GSs. Here, the red and black dashed curves refer
to unstable solitons which evolve, respectively, into robust breathers [see
Fig.\protect\ref{offoddanti}(c)], or into a confined turbulent state shown
in Fig.\protect\ref{offoddanti}(d). (c) The detailed structure of the
red-dotted segment with alternate stability and instability in panel (a).
The red triangles and blue filled circles label stable and weakly unstable
solutions, respectively.}
\label{offsite:stability}
\end{figure}

For antisymmetric states, with $U(x)=-V(x)$ (in the system with $b=0$), the
above-mentioned effective single-component lattice potential, generated by
the Rabi lattice, inverts its sign, taking the form of $\epsilon \cos \left(
2x\right) $. Accordingly, off-site-centered antisymmetric GSs tend to
minimize their energy by placing two density maxima at potential-minima
points, $x=\pm \pi /2$, which gives such solitons a chance to be stable. A
typical example of a stable antisymmetric GS with the off-site-centered
spatially-even shape is shown in Fig. \ref{offevenanti}. The stability of
this GS species is summarized in Fig. \ref{offsite:stability}(a), which
demonstrates that they are stable solely in the first finite bandgap [the
instability in the second finite bandgap is illustrated by Fig. \ref%
{offevenanti}(d)]. Similar to the situation shown for the on-site-centered
symmetric GSs in Fig. \ref{Syrelationship}(b), at the right edge of the
bandgap there is a narrow segment of alternating stability and weak
oscillatory instability, whose structure is displayed in detail in Fig. \ref%
{offsite:stability}(c).

\begin{figure}[tbp]
\centering{\label{fig8a} \includegraphics[scale=0.2]{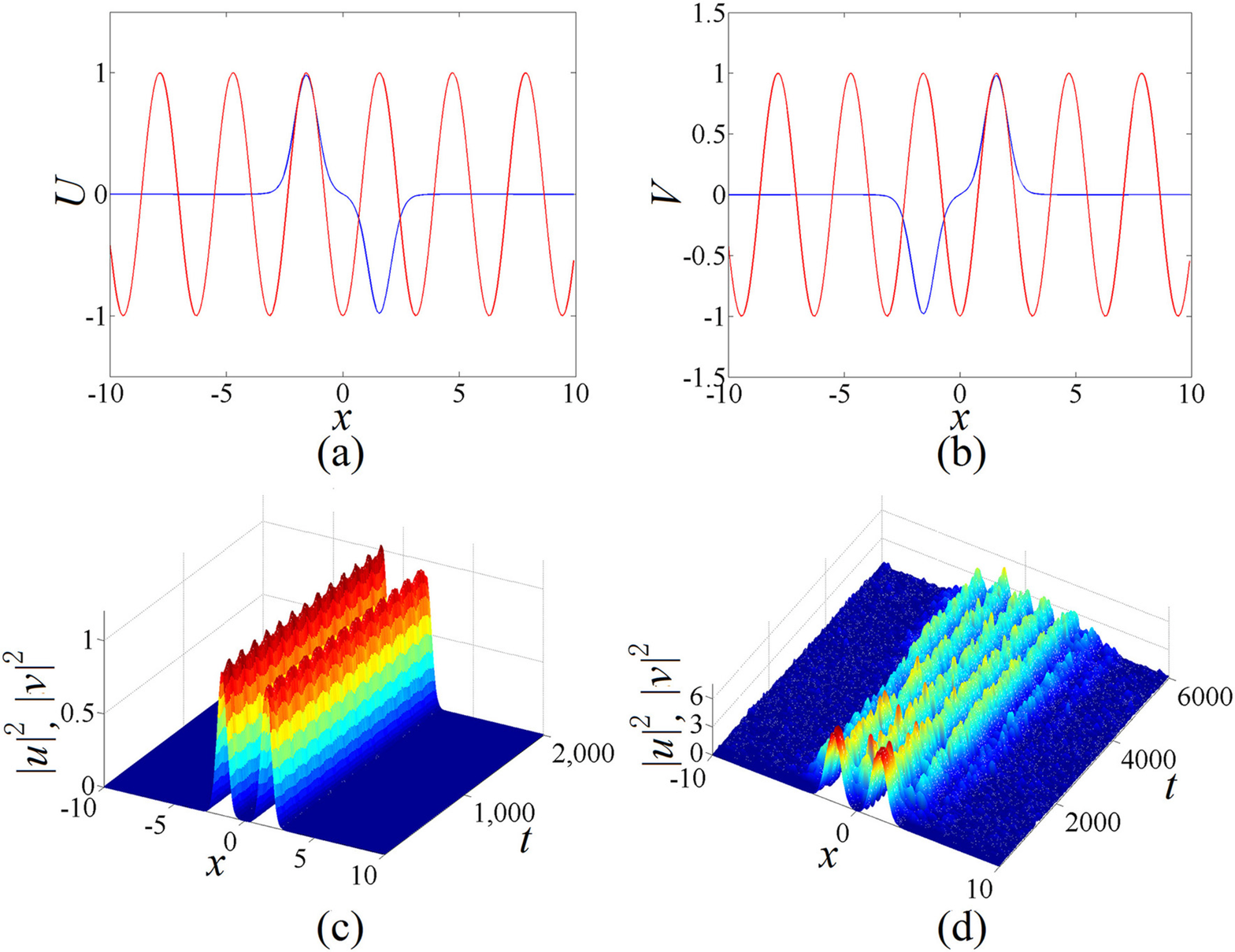}}
\caption{(Color online) (a,b) An example of an unstable off-site-centered
spatially-odd antisymmetric GS for $\protect\mu =-3$ (which belongs to the
first finite bandgap) with $b=0$ and $\protect\epsilon =6$. (c) The
perturbed evolution (with initial random perturbations at the $5\%$ level)
of the same soliton shows its transformation into a persistent breather. (d)
Simulations of the perturbed evolution of the GS of the same type, but with $%
\protect\mu =2$ (which falls into the second finite bandgap) show its
transformation into a confined turbulent mode.}
\label{offoddanti}
\end{figure}

\subsection{Other types of symmetric and antisymmetric gap solitons}

The system with $b=0$ supports off-site-centered antisymmetric GSs with the
spatially-odd shape, in addition to their even counterparts considered
above. An example, and the family of such solitons in the $(\mu ,N)$ plane,
are displayed in Figs. \ref{offoddanti} and \ref{offsite:stability}(b),
respectively. The conclusion of the analysis is that the family is
completely unstable in both first and second finite bandgaps. Further,
direct simulations demonstrate that, in the first bandgap, the instability
transforms the stationary spatially-odd GSs into persistent breathers, see
Fig. \ref{offoddanti}(c), while in the second bandgap, the unstable GSs
evolve into apparently turbulent spatially confined modes, as shown in Fig. %
\ref{offoddanti}(d). In either case, the dynamical states produced by the
instability keep the spatially odd shape, as indicated by the persistence of
zero amplitude at the midpoint in Figs. \ref{offoddanti}(c) and \ref%
{offoddanti}(d).

Additional types of solitons
%, which are displayed in Fig. \ref{threetypes},
have been found too: on-site-centered antisymmetric, off-site-centered
spatially-even symmetric, and off-site-centered spatially-odd symmetric GSs.
Due to the nature of the present system, in which the GSs are supported by
the Rabi lattice, these additional species are actually tantamount to the
three species considered above. Indeed, the on-site-centered symmetric GS is
obviously equivalent to an antisymmetric one which is centered at the site
shifted by half a spatial period, %cf. Figs. \ref{threetypes}(a1,a2) and %
%\ref{systable},
their stability being identical too. The same pertains to off-site-centered
spatially-even and odd symmetric GSs, which may be easily converted into
their antisymmetric counterparts. %{\LARGE I have
%commented out former Fig. 9, as it occupied a lot of space, but did not show
%anything new, as a matter if fact.}

%\begin{figure}[tbp]
%\centering{\label{fig9a} \includegraphics[scale=0.17]{9a.eps}}
%\caption{(Color online) (a1, a2), (b1, b2) and (c1, c2) are examples of
%on-site-centered antisymmetric, off-site-centered spatially-even symmetric,
%and off-site-centered spatially-odd symmetric GSs, respectively. They all
%were found in the first finite bandgap, with $\protect\mu =-2$. Here, the
%first two solitons are stable, while the last one is unstable, which evolves
%into robust breather. }
%\label{threetypes}
%\end{figure}

\section{On-site-centered gap solitons in the asymmetric system}

\subsection{Numerical results}

It is relevant to stress that, while GSs in symmetric dual-core systems with
the usual lattice potential and constant inter-core coupling readily feature
spontaneous breaking of the (anti-)symmetry between their components,
followed by generation of asymmetric solitons, provided that the
nonlinearity strength exceeds a certain critical value \cite%
{Gubeskys,Matuszewski}, this effect is not observed in the present system,
i.e., all the GSs, both stable and unstable ones, are either symmetric or
antisymmetric with respect to the two components. In this section, we report
results for the most fundamental on-site-centered asymmetric GSs, which are
naturally produced by the asymmetric system, based on Eqs. (\ref{basicEqU}),
(\ref{basicEqV}) with $b\neq 0$ (obviously, it is sufficient to consider $%
b>0 $). We do not aim here to consider other GS species in the asymmetric
system.

\begin{figure}[tbp]
\centering{\label{fig9a} \includegraphics[scale=0.2]{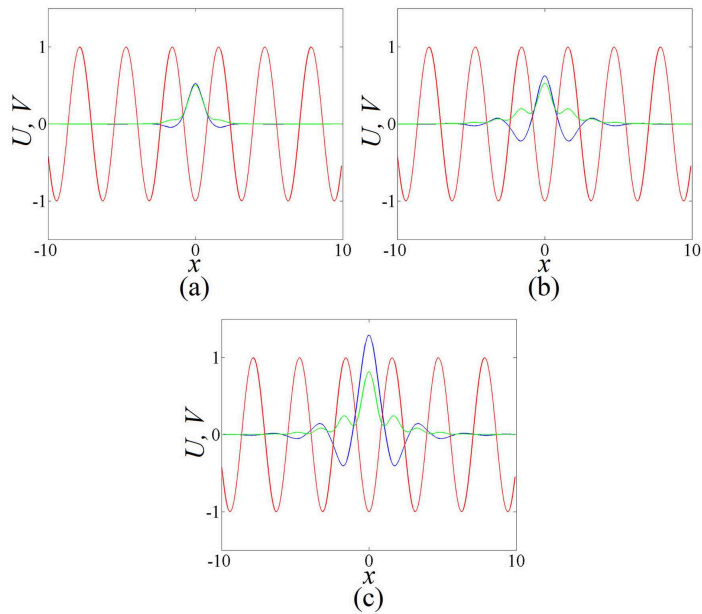}}
\caption{(Color online) Examples of stable on-site-centered asymmetric GSs
with $\protect\epsilon =6$, and fixed chemical potential $\protect\mu =-3.5$%
: (a) $b=0.2$; (b) $b=1$; (c) $b=3$. Fields $U(x)$ and $V(x)$ with larger
and smaller amplitudes are shown by blue and green lines, respectively.}
\label{asypatern}
\end{figure}

\begin{figure}[tbp]
\centering{\label{fig10a} \includegraphics[scale=0.2]{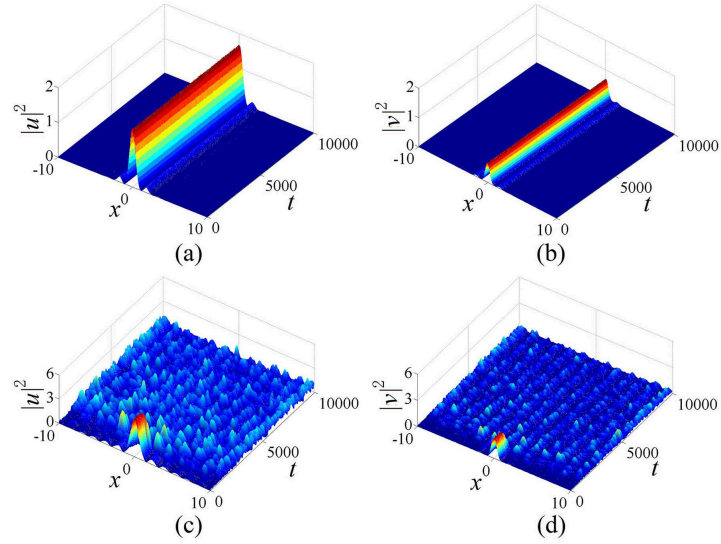}}
\caption{(Color online) (a,b) Direct simulations of the perturbed evolution
of a stable asymmetric on-site-centered GSs with $(b,\protect\mu )=(3,-3.5)$%
. (c,d) The same for an unstable asymmetric soliton, with $(b,\protect\mu %
)=(3,-1.5)$.}
\label{asysimulation}
\end{figure}

Examples of stable asymmetric GSs are presented in Fig. \ref{asypatern}.
With the increase of $b$, the shapes of the $U(x)$ and $V(x)$ components
become less localized and develop undulations in their tails. Examples of
the perturbed evolution of stable and unstable asymmetric GSs are further
shown in Fig. \ref{asysimulation}. Different from the case of symmetric
on-site-centered GSs, asymmetric ones which are unstable tend to develop a
chaotic state expanding to the entire spatial domain, cf. Fig. \ref%
{syunstable}(c,d)

\begin{figure}[tbp]
\centering{\label{fig11a} \includegraphics[scale=0.15]{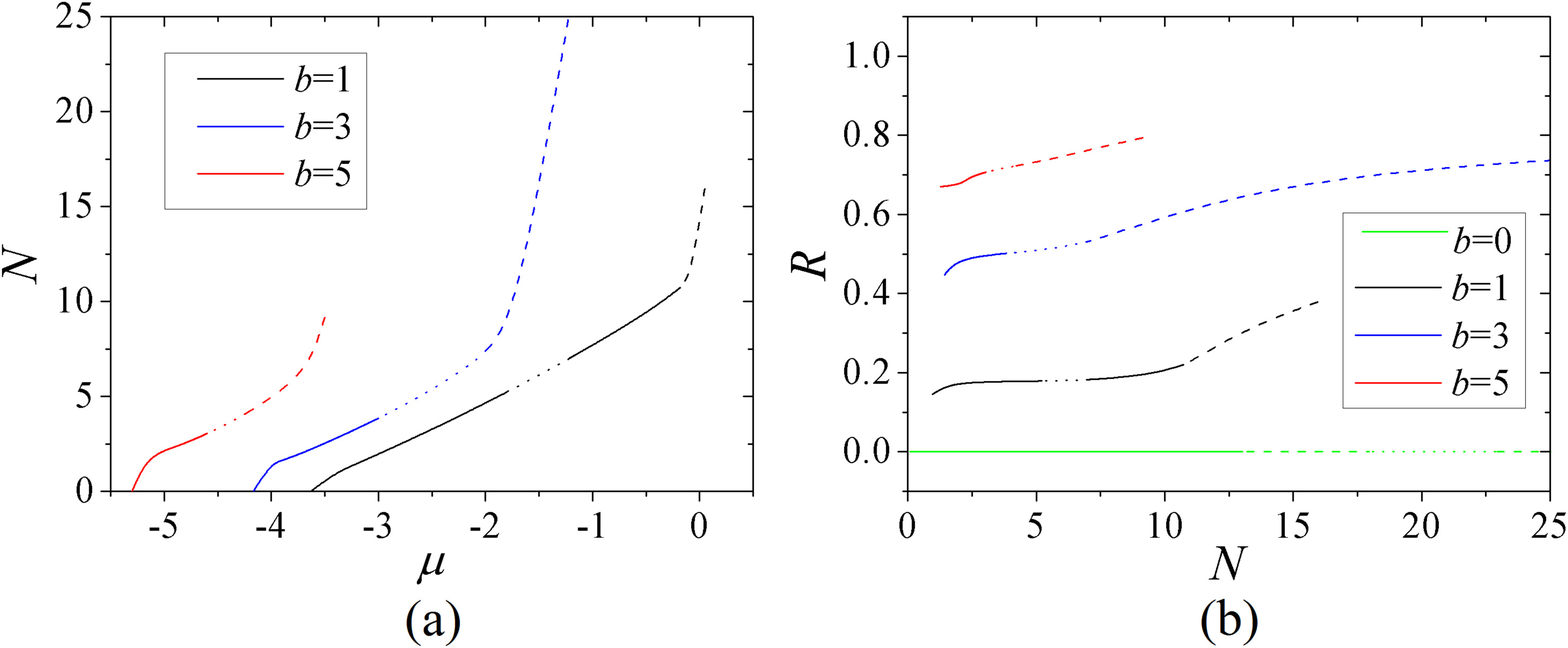}}
\caption{(Color online) (a) Total norm $N$ of asymmetric on-site-centered
GSs versus chemical potential $\protect\mu $, at $\protect\epsilon =6$ and
fixed values of the asymmetry coefficient: $b=1,b=3$, and $b=5$. (b)
Asymmetry ratio, $R$, defined as per Eq. (\protect\ref{R}), versus $N$, for
the same soliton families. The family with $b=0$, which has $R\equiv 0$, is
included too, for the completeness' sake. In these panels, solid and dashed
lines represent stable and unstable GSs, respectively, while the dotted
segments designate regions of alternate stability and instability.}
\label{asyregion}
\end{figure}

\begin{figure}[tbp]
\centering{\label{fig12a} \includegraphics[scale=0.15]{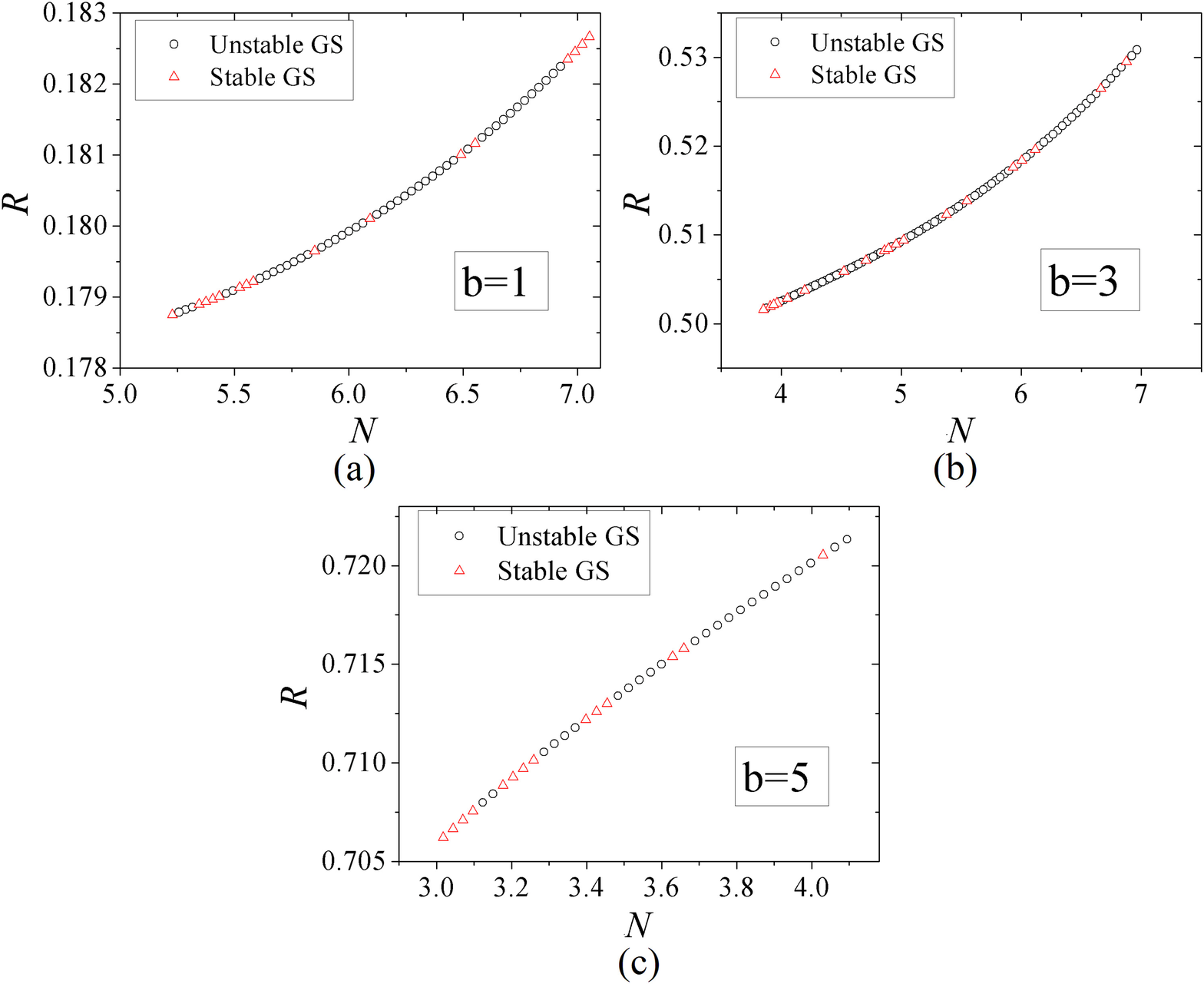}}
\caption{(Color online) Panels (a), (b), and (c) show the detailed structure
of regions with alternating stable and unstable on-site-centered asymmetric
GSs in the regions shown by the dotted line in Fig. \protect\ref{asyregion}%
(a), for $b=1,3,$ and $5$, respectively. Red triangles and black circles
represent, severally, stable and unstable solitons.}
\label{asyalter}
\end{figure}

Results obtained for families of asymmetric GSs and their stability are
summarized in Fig. \ref{asyregion}, by means of $N(\mu )$ and $R(N)$ curves
[recall $R$ is the asymmetry ratio defined in Eq. (\ref{R})]. These
dependences suggest that the asymmetric GSs are more stable at lower values
of the total norm, $N$, and stability regions shrink with the increase of
asymmetry coefficient $b$ and asymmetry ratio $R$. Similar to the symmetric
system ($b=0$), here there also exist regions of alternate stability and
instability, which are designated by dotted segments in Fig. \ref{asyregion}%
. The detailed structure of these segments is rendered in Fig. \ref{asyalter}%
.

\begin{figure}[tbp]
\centering{\label{fig13a} \includegraphics[scale=0.15]{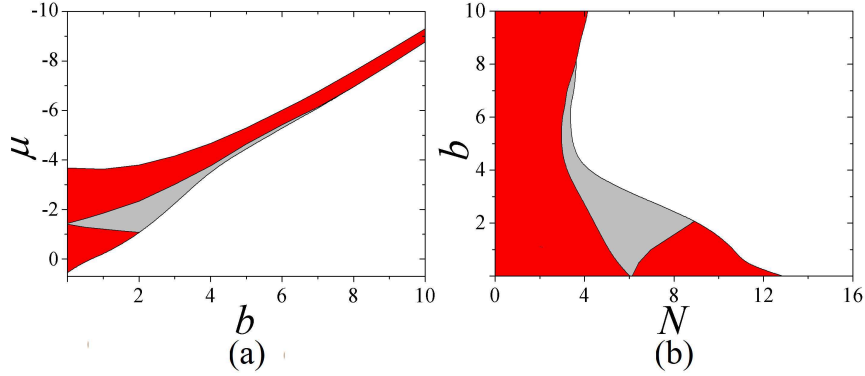}}
\caption{(Color online) Stability diagram for on-site-centered asymmetric
GSs in the planes of $(b,\protect\mu )$ (a) and $(N,b)$ (b). The red and
gray colors designate, respectively, the stability area, and the one of
alternate stability and instability. GSs do not exist in the top blank areas
in panels (a). In the bottom blank area in (a) and blank area in (b), there
exists completely unstable GSs.}
\label{stabilityarea}
\end{figure}

In Fig. \ref{stabilityarea}, we have collected results produced by the
stability analysis for the on-site asymmetric GSs in the planes of $(b,\mu )$
and $(N,b)$. Figures \ref{stabilityarea}(a,b) demonstrate that the stability
area originally shrinks with the increase of the asymmetry coefficient, $b$,
and then stays narrow but nearly constant. In the $(N,b)$ plane, the
stability region also narrows at first with the increase of $b$, but then it
broadens at still larger $b$.

\subsection{A semi-analytical approximation}

In the limit of $b\gg 1$, the two-component system, based on Eqs. (\ref%
{basicEqU}) and (\ref{basicEqV}) can be easily reduced to a single equation
with the usual lattice potential, by means of an approximation similar to
that recently elaborated for the two-component BEC\ under the action of
strong Zeeman splitting in Ref. \cite{Sherman}. The approximation is based
on the fact that large $b$ gives rise to solutions with chemical potential $%
\mu =-b+\delta \mu ,$ which implies $\left\vert \delta \mu \right\vert \ll b$%
. Accordingly, the solutions are looked for in the form of $\left\{ u\left(
x,t\right) ,v\left( x,t\right) \right\} =\exp \left( ibt\right) \left\{
\tilde{u}\left( x,t\right) ,\tilde{v}\left( x,t\right) \right\} $, where the
remaining $t$-dependence in $\tilde{u}$ and $\tilde{v}$ is slow, in
comparison with $\exp (ibt)$. Then Eq. (\ref{basicEqV}), in which the the
XPM terms may be restored, as per Eq. (\ref{g}), readily yields an
approximate expression for the small $v$ component:
\begin{equation}
\tilde{v}\approx \frac{\epsilon \cos \left( 2x\right) }{2b+\sigma g|u|^{2}}%
\tilde{u}\approx \epsilon \left( \frac{1}{2b}-\frac{\sigma g}{4b^{2}}%
|u|^{2}\right) \cos (2x)\cdot \tilde{u}.  \label{vu}
\end{equation}%
To justify the expansion of the the fraction in this expression, it is
assumed that the soliton's peak density is not too large, \textit{viz}.,%
\begin{equation}
\left( |u|^{2}\right) _{\max }\ll b/|g|  \label{<<}
\end{equation}%
(this condition is easily satisfied for large $b$). Then, the substitution
of this approximation in Eq. (\ref{basicEqU}) leads to a single equation for
$\tilde{u}\left( x,t\right) $, with an effective lattice potential:%
\begin{equation}
\left[ i{\frac{\partial }{\partial t}}+{\frac{1}{2}}{\frac{\partial ^{2}}{%
\partial x^{2}}}-\sigma \left( 1+\frac{\epsilon ^{2}g}{8b^{2}}\right) |%
\tilde{u}|^{2}+\frac{\epsilon ^{2}}{2b}\cos ^{2}(2x)\right] \tilde{u}=0
\label{delta_k}
\end{equation}%
[it is easy to see that, to obtain the nonlinearity coefficient in Eq. (\ref%
{delta_k}) under the above condition (\ref{<<}), one may substitute $\cos
^{2}(2x)$ by its average value, $1/2,$ while this substitution is not
relevant in the effective lattice potential]. In turn, Eq. (\ref{delta_k})
with $\sigma \left( 1+\epsilon ^{2}g/8b^{2}\right) >0$ is the standard
equation which gives rise to the usual GSs \cite{DPeli,Yang}. Note that the
renormalization of the nonlinearity coefficient in Eq. (\ref{delta_k}),
represented by the term proportion to the XPM coefficient, $g$, is essential
if the Rabi lattice is strong enough, namely, $|g|\epsilon ^{2}\sim 8b^{2}$,
which is compatible with the underlying condition $b\gg 1$.

\begin{figure}[tbp]
\centering{\label{fig14a} \includegraphics[scale=0.2]{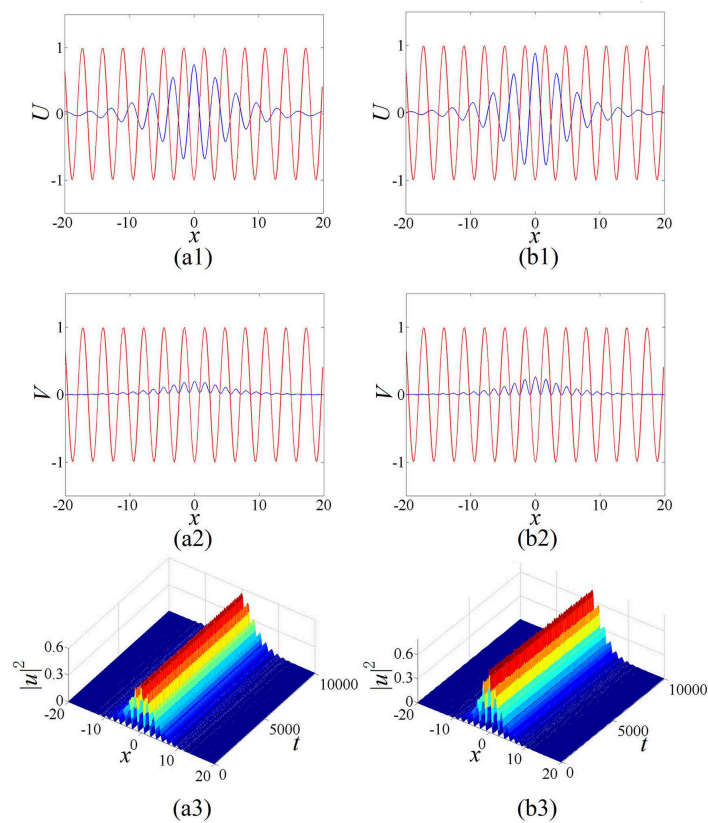}}
\caption{(Color online) Panels (a1,a2) and (b1,b2) present typical examples
of stable GSs produced, respectively, by the full system of Eqs. (\protect
\ref{basicEqU}) and (\protect\ref{basicEqV}), and by the semi-analytical
approximation which amounts to the single equation (\protect\ref{delta_k}),
for $(b,\protect\mu )=(10,-9.1)$. Results of direct simulations, initiated
by the full two-component solution, and by the single-component
approximation, are displayed, for the $u$ component, in panels (a3) and
(b3), respectively.}
\label{comparison}
\end{figure}

This approximation may be naturally named semi-analytical, as its analytical
part replaces the underlying system of two Gross-Pitaevskii equations,
coupled by the Rabi lattice, by the single standard equation (\ref{delta_k}%
), whose solution is not, generally, available analytically, but is known
very well in the numerical form \cite{DPeli,Yang,Alfimov}. Comparison
between a typical stationary GS predicted by this approximation and its
counterpart produced by the numerical solution of Eqs. (\ref{statEqsU}) and (%
\ref{statEqsV})\ is presented in Fig. \ref{comparison}. Obviously, the
semi-analytical and numerical profiles are close to each other, and both are
stable. In fact, the broad modes displayed in this figure resemble the known
nonlinear states generalizing GSs in the usual single-component model, in
the form of ``truncated Bloch waves", which were
demonstrated experimentally in Ref. \cite{Oberthaler} and explained
theoretically in Ref. \cite{Canberra}.

%{\LARGE [I have made a chance in the caption to Fig. \ref{comparison}. Pleae
%check it.]}

\section{Conclusion}

The objective of this work is to introduce the model of two-component gap
solitons, based on two GPEs (Gross-Pitaevskii equations) with self-repulsive
nonlinearity, coupled by linear terms which are subject to the spatially
periodic cosinusoidal modulation (\textit{Rabi lattice}). The system can be
implemented in binary BEC with the superimposed standing wave of a
resonantly-coupling electromagnetic field, and in the bimodal light
propagation in twisted fibers. We have demonstrated that this setting gives
rise to stable two-component GSs (gap solitons), in the absence of periodic
potentials which are necessary for the existence of GSs in usual models.
Several types of the GSs have been found, including on-site-centered
symmetric and antisymmetric modes, and spatially even and odd
off-site-centered symmetric and antisymmetric ones. Both symmetric and
antisymmetric GSs are stable chiefly in the first finite bandgap, as well as
in a small segment of the second bandgap. These findings are, roughly,
similar to what was recently found for the stability of usual GSs in the
single-component model with a periodic potential \cite{Alfimov}. A
noteworthy finding is the alternation of stable and unstable
on-site-centered symmetric GSs in the latter segment. For on-site-centered
symmetric and off-site-centered spatially even symmetric and antisymmetric
GSs, there also exists a narrow segment of alternating stability and weak
oscillatory instability near the right edge of first finite bandgap.
Unstable GSs spontaneously transform into robust breathers or spatially
confined turbulent states. On-site-centered GSs were found in the asymmetric
system too, where segments featuring the alternate stability exist as well.
Thus, the alternation of stability and instability of GSs, which was not
reported in previously studied models, is a characteristic generic feature
of the present system. It is worthy to note that the stability area of the
on-site-centered asymmetric GSs originally shrinks with the increase of the
asymmetry coefficient, $b$, but then it expands with the further increase of
$b$, in terms of the total power of the solitons. In the limit of $b\gg 1$,
an analytical approximation makes it possible to transform the system into a
single GPE with an effective periodic potential and respective GS solutions.

A natural extension of the present analysis should produce a detailed
analysis of the system including the XPM interaction between the two
components. A challenging direction for further work is a two-dimensional
version of the present system. In that case, two-component solitary vortices
may be looked for, in addition to fundamental GSs. In fact,
off-site-centered spatially odd GSs, which are considered above, are
one-dimensional counterparts of the two-dimensional vortices.

\section*{Acknowledgments}

This work was supported, in part, by grant No. 2015616 from the joint
program in physics between NSF and Binational (US-Israel) Science
Foundation. We appreciate valuable discussions with G. L. Alfimov.

\bibliographystyle{plain}
\bibliography{apssamp}

\begin{thebibliography}{99}
\bibitem{early-reviews} V. A. Brazhnyi and V. V. Konotop, Mod. Phys. Lett. B
\textbf{18}, 627 (2004); O. Morsch and M. Oberthaler, Rev. Mod. Phys.
\textbf{78}, 179 (2006).

\bibitem{DPeli} D. E. Pelinovsky, \textit{\ Localization in Periodic
Potentials: from Schr\"{o}dinger Operators to the Gross-Pitaevskii Equation}
(Cambridge University Press: Cambridge, 2011); B. A. Malomed, D. Mihalache,
F. Wise, and L. Torner, \textit{Spatiotemporal optical solitons}, J. Optics
B: Quant. Semicl. Opt. \textbf{7}, R53-R72 (2005).

\bibitem{Yang} J. Yang, \textit{Nonlinear Waves in Integrable and
Nonintegrable Systems}. (SIAM: Philadelphia, 2010).

\bibitem{Kartashov} Y. V. Kartashov, B. A. Malomed, and L. Torner, Rev. Mod.
Phys. \textbf{83}, 247-306 (2011).

\bibitem{Bragg} C. M. de Sterke and J. E. Sipe, Progr. Optics \textbf{33},
203 (1994); C. R. Giles, J. Lightwave Tech. \textbf{15}, 1391 (1997).

\bibitem{Lederer} F. Lederer, G. I. Stegeman, D. N. Christodoulides, G.
Assanto, M. Segev, and Y. Silberberg, Phys. Rep. \textbf{463}, 1 (2008).

\bibitem{Joannopoulos} J. D. Joannopoulos, S. G. Johnson, J. N. Winn, and R.
D. Meade, \textit{Photonic Crystals: Molding the Flow of Light} (Princeton
University Press: Princeton, 2008).

\bibitem{Szameit} A. Szameit, D. Bl\"{o}mer, J. Burghoff, T. Schreiber, T.
Pertsch, S. Nolte, and A. T\"{u}nnermann, Opt. Express \textbf{13},
10552-10557(2005).

\bibitem{Efremidis} N. K. Efremidis, S. Sears, D. N. Christodoulides, J. W.
Fleischer, and M. Segev, Phys. Rev. E \textbf{66}, 046602 (2002); J. W.
Fleischer, G. Bartal, O. Cohen, T. Schwartz, O. Manela, B. Freedman, M.
Segev, H. Buljan, and N. K. Efremidis, Opt. Exp. \textbf{13}, 1780-1796
(2005).

\bibitem{Pitaevskii} L. Pitaevskii and S. Stringari, \textit{Bose-Einstein
Condensate} (Clarendon Press: Oxford, 2003); H. T. C. Stoof, K. B. Gubbels,
and D. B. M. Dickerscheid, \textit{Ultracold Quantum Fields} (Springer:
Dordrecht, 2009).

\bibitem{Wu} B. Wu and Q. Niu, Phys. Rev. A \textbf{64}, 061603(R) (2001).

\bibitem{Salerno} V. V. Konotop, M. Salerno, Phys. Rev. A \textbf{65},
021602 (2002); G. L. Alfimov, V. V. Konotop and M. Salerno, Europhys. Lett.
\textbf{58}, 7-13 (2002); G. L. Alfimov, P. G. Kevrekidis, V. V. Konotop,\
and M. Salerno, Phys. Rev. E \textbf{66}, 046608 (2002).

\bibitem{Oberthaller} K.M. Hilligsoe, M. K. Oberthaler, and K. P. Marzlin,
Phys. Rev. A \textbf{66}, 063605 (2002); B. Eiermann, Th. Anker, M. Albiez,
M. Taglieber, P. Treutlein, K.-P. Marzlin, and M. K. Oberthaler, Phys. Rev.
Lett. \textbf{92,} 230401 (2004).

\bibitem{Louis} P. J. Y. Louis, E. A. Ostrovskaya, C. M. Savage and Yu. S.
Kivshar, Phys. Rev. A \textbf{67}, 013602 (2003); D. E. Pelinovsky, A. A.
Sukhorukov and Yu. S. Kivshar, Phys. Rev. E \textbf{70}, 036618 (2004).

\bibitem{HS} H. Sakaguchi and B. A. Malomed, J. Phys. B \textbf{37},
1443-1459 (2004).

\bibitem{Myatt} C. J. Myatt, E. A. Burt, R. W. Ghrist, E. A. Cornell, and C.
E. Wieman, Phys. Rev. Lett. \textbf{78}, 586 (1997); D. M. Stamper-Kurn, M.
R. Andrews, A. P. Chikkatur, S. Inouye, H.-J. Miesner, J. Stenger, and W.
Ketterle, \textit{ibid}. \textbf{80}, 2027 (1998).

\bibitem{Mak} W. C. K. Mak, B. A. Malomed, and P. L. Chu, J. Opt. Soc. Am. B
\textbf{15}, 1685-1692 (1998).

\bibitem{Harel} A. Harel and B. A. Malomed, Phys. Rev. A \textbf{89}, 043809
(2014).

\bibitem{Skryabin} A. V. Yulin, D. V. Skryabin, and W. J. Firth, Phys. Rev.
E \textbf{66}, 046603 (2002).

\bibitem{Gubeskys} A. Gubeskys, B. A. Malomed, and I. M. Merhasin, Phys.
Rev. A \textbf{73}, 023607 (2006); A. Gubeskys and B. A. Malomed, \textit{%
ibid}. \textbf{75}, 063602 (2007).

\bibitem{Matuszewski} M. Matuszewski, B. A. Malomed, and M. Trippenbach ,
Phys. Rev. A \textbf{76}, 043826 (2007); M. Trippenbach, E. Infeld, J.
Gocalek, M. Matuszewski, M. Oberthaler, and B. A. Malomed, \textit{ibid}.
\textbf{78}, 013603 (2008).

\bibitem{Adhikari} S. K. Adhikari and B. A. Malomed, Phys. Rev. A \textbf{77}%
, 023607 (2008); \textit{ibid}. A \textbf{79}, 015602 (2009).

\bibitem{Athikom} A. Roeksabutr, T. Mayteevarunyoo, and B. A. Malomed. Opt.
Exp. \textbf{20}, 24559-24574 (2012).

\bibitem{dark} H. Ter\c{c}as, D. D. Solnyshkov, and G. Malpuech, Phys. Rev.
Lett. \textbf{110}, 035303 (2013); \textit{ibid}. \textbf{113}, 036403
(2014).

\bibitem{Perez} V. M. P\'{e}rez-Garc\'{\i}a and J. B. Beitia, Phys. Rev. A
\textbf{72}, 033620 (2005); S. K. Adhikari, \textit{ibid}. \textbf{70},
043617 (2004); \textbf{72},053608 (2005); \textbf{76}, 053609 (2007); Phys.
Lett. A \textbf{346}, 179 (2005); J. Phys. A: Math. Theor. \textbf{40}, 2673
(2007).

\bibitem{Simoni} S. Inouye, M. R. Andrews, J. Stenger, H. J. Miesner, D. M.
Stamper-Kurn, and W. Ketterle, Nature (London) \textbf{392}, 151 (1998); A.
Simoni, F. Ferlaino, G. Roati, G. Modugno and M. Inguscio, Phys. Rev. Lett.
\textbf{90}, 163202 (2003); M. Theis, G. Thalhammer, K. Winkler, M. Hellwig,
G. Ruff, R. Grimm, and J. H. Denschlag, Phys. Rev. Lett. \textbf{93}, 123001
(2004)..

\bibitem{Ballagh} R. J. Ballagh, K. Burnett, and T. F. Scott, Phys. Rev.
Lett. \textbf{78}, 1607 (1997); J. Williams, R. Walser, J. Cooper, E.
Cornell, M. Holland, Phys. Rev. A \textbf{59}, R31 (1999); P. \"{O}hberg and
S. Stenholm, \textit{ibid}. \textbf{59}, 3890 (1999); D. T. Son and M. A.
Stephanov, \textit{ibid}. \textbf{65}, 063621 (2002); S. D. Jenkins and T.
A. B. Kennedy, \textit{ibid}. \textbf{68}, 053607 (2003); H. Saito, R. G.
Hulet, and M. Ueda, \textit{ibid}. \textbf{76}, 053619 (2007); H. Guo, Z.
Chen, J. Liu and Y. Li, Laser Phys. \textbf{24}, 045403 (2014); J. Qin, G.
Dong, and B. A. Malomed, Phys. Rev. Lett. \textbf{115}, 023901 (2015).

\bibitem{Rabi} H. Susanto, P. G. Kevrekidis, B. A. Malomed, and F. Kh.
Abdullaev, Phys. Lett. A \textbf{372}, 1631 (2008).

\bibitem{Yongyao1} Y. Li, W. Pang, S. Fu, and B. A. Malomed, Phys. Rev. A
\textbf{85}, 053821 (2012).

\bibitem{coupler} A. W. Snyder, D. J. Mitchell, L. Poladian, D. R. Rowland,
and Y. Chen, J. Opt. Soc. Am. B \textbf{8}, 2102 (1991); M. Romangoli, S.
Trillo, and S. Wabnitz, \textbf{24}, S1237 (1992); W.-P. Huang, J. Opt. Soc.
Am. A \textbf{11}, 963 (1994).

\bibitem{BEC-coupler} A. Smerzi, S. Fantoni, S. Giovanazzi, S. R. Shenoy,
Phys. Rev. Lett. \textbf{79}, 4950-4953 (1997).

\bibitem{Zlattice} Y. V. Kartashov, V. V. Konotop, and F. Kh. Abdullaev,
Phys. Rev. Lett. \textbf{111}, 060402 (2013).

\bibitem{cross-F} S. B. Papp, J. M. Pino, and C. E. Wieman, Phys. Rev. Lett.
\textbf{101}, 040402 (2008); F. Wang, X. Li, D. Xiong, and D. Wang, J. Phys.
B \textbf{49}, 015302 (2016).

\bibitem{rocking} S. Trillo, S. Wabnitz, W. C. Banyai, N. Finlayson, C. T.
Seaton, and G. I, Stegeman, IEEE J. Quant. Elect. \textbf{25}, 101 (1989).

\bibitem{PCF} R. Salem, A. S. Lenihan, G. M. Carter, T. E. Murphy, IEEE J.
Sel. Top. Quant. Elect. \textbf{14}, 540 (2008); L. Y. Zang, M. S. Kang, M.
Kolesik, M. Scharrer, and P. Russell, J. Opt. Soc. Am. B \textbf{27}, 1742
(2010).

\bibitem{Alfimov} G. L. Alfimov, P. P. Kizin, and D. A. Zezyulin, Physica D
\textbf{337}, 58 (2016).

\bibitem{Sherman} H. Sakaguchi, E. Ya. Sherman, and B. A. Malomed, Phys.
Rev. E \textbf{94}, 032202 (2016).

\bibitem{Oberthaler} T. Anker, M. Albiez, R. Gati, S. Hunsmann, B. Eiermann,
A. Trombettoni, and M. K. Oberthaler, Phys. Rev. Lett. \textbf{94}, 020403
(2005).

\bibitem{Canberra} J. Wang, Jianke Yang, T. J. Alexander, and Y. S. Kivshar,
Phys. Rev. A \textbf{79}, 043610 (2009).
\end{thebibliography}
% Produces the bibliography via BibTeX.

\end{document}